\documentclass{pas}

\usepackage{aas-macros}
\usepackage{multirow}

\newcommand{\beq}{\begin{equation}}
\newcommand{\eeq}{\end{equation}}
\newcommand{\beqa}{\begin{eqnarray}}
\newcommand{\eeqa}{\end{eqnarray}}

\begin{document}

\lefttitle{Publications of the Astronomical Society of Australia}
\righttitle{Cambridge Author}

\jnlPage{1}{4}
\jnlDoiYr{2025}
\doival{10.1017/pasa.xxxx.xx}

\articletitt{Research Paper}

\title{The effect of collisional cooling of energetic electrons on radio emission from  the centrifugal magnetospheres of magnetic hot stars}

\author{\sn{B.} \gn{Das}$^{1}$ and \sn{S. P.} \gn{Owocki}$^{2}$}

\affil{$^1$CSIRO, Space and Astronomy, P.O. Box 1130, Bentley WA 6102, Australia}
\affil{$^2$Department of Physics and Astronomy, University of Delaware, 217 Sharp Lab, Newark, Delaware, 19716, USA}

\corresp{B. Das, Email: Barnali.Das@csiro.au}



\begin{abstract}
This paper extends our previous study of the gyro-emission by energetic electrons in the magnetospheres of rapidly rotating, magnetic massive stars,
through a quantitative analysis of the role of cooling by Coulomb collisions with thermal electrons from stellar wind material trapped within the centrifugal magnetosphere (CM).
For the standard, simple CM model of a dipole field with aligned magnetic and rotational axes, we show that 
both gyro-cooling along magnetic loops and Coulomb cooling in the CM layer have nearly the same dependence on the magnitude and radial variation of magnetic field, implying then that their ratio is a {\it global} parameter that is largely {\it independent} of the field.
Analytic analysis shows that, for electrons introduced near the CM layer around a magnetic loop apex, collisional cooling is more important
for electrons with high pitch angle, while more field-aligned electrons cool by gyro-emission near their mirror point close to the loop base.
Numerical models that assume a gyrotropic initial deposition with a gaussian distribution in both radius and loop co-latitude show the 
residual gyro-emission is generally strongest near the loop base, with highly relativistic  electrons suffering much lower collisional losses than lower-energy electrons that are only mildly relativistic.
Even for cases in which the energy deposition is narrowly concentrated near the loop apex,
the computed residual emission shows a surprisingly broad distribution with magnetic field strength,
suggesting that associated observed radio spectra should generally have a similarly broad frequency distribution.
Finally, we briefly discuss the potential applicability of this formalism to magnetic ultracool dwarfs (UCDs), for which VLBI observations indicate incoherent radio emission to be concentrated around the magnetic equator, in contrast to our predictions here for magnetic hot stars. We suggest that this difference could be attributed to UCDs having either a lower ambient density of thermal electrons, or more highly relativistic non-thermal electrons, both of which would reduce the relative importance of the collisional cooling explored here.
\end{abstract}

\begin{keywords}
stars: magnetic field -- stars: early type -- stars: rotation -- radio continuum: stars -- magnetic reconnection
\end{keywords}

\maketitle

\section{Introduction}

Hot magnetic stars with moderately rapid rotation show incoherent, circularly polarized radio emission, thought to arise from gyro-synchrotron emission of energetic electrons trapped in magnetic loops
\citep[e.g.][]{drake1987,andre1988,trigilio2004}.
Empirical analyses show that the radio emission depends on  both the magnetic field strength and the stellar rotation rate
\citep{leto2021,shultz2021},
with a scaling that is well explained by a model in which the electrons are energized by magnetic reconnection
events that arise from {\it centrifugal breakout} (CBO) of plasma trapped in the rotating magnetosphere \citep{owocki2022}.

A previous paper \citep[][hereafter Paper\,I]{Das2023} examined how gyro-synchrotron radio emission cools the electrons, and how this affects the spatial and spectral distribution of the observed radio emission.
Paper\,I provided a broader discussion of the history of observational and theoretical studies of such incoherent radio emission from these magnetic hot stars, and also presented a preliminary initial discussion of the potential importance of additional cooling from Coulomb collisions with ambient thermal electrons.

Building on this, the present paper now carries out a detailed analysis of such collisional cooling by electrons in the dense  ``centrifugal magnetosphere" (CM) layer that forms in the common rotational and magnetic equator of rapidly rotating early-type stars with a rotation-aligned dipole field\footnote{The generalization to oblique dipoles is left to future study.} \citep{petit2013,owocki2020}.
As in Paper\,I, the analysis here assumes that repeated CBO-driven magnetic reconnection events seed a quasi-steady, gyrotropic population of energetic electrons around the tops of underlying closed magnetic loops.
In the standard scenario \citep[e.g.][]{leto2021}, the spiraling of these energetic electrons as they mirror between the opposite footpoints of the loop leads to the gyro-synchrotron emission of the observed radio.

The present study now accounts (Section \ref{subsec:cooling_CM}) for the Coulomb collisional energy loss of such mirroring energetic electrons each time they cross the dense CM layer near the loop apex (which here lies in the common rotational/magnetic equator), showing how this scales with electron Lorentz factor $\gamma$ and stellar parameters (see Equation\ (\ref{eq:tauc1})).
Comparison (Section \ref{subsec:reduced_gyrocool}) of such CM-cooling and gyro-cooling shows they each have nearly the same dependencies on the magnetic field strength and its radial variation, so that their ratio becomes a global parameter that is surprisingly {\it independent} of the field; but its overall value (see Equations\ (\ref{eq:kbtauc}) and (\ref{eq:Ckdef})) confirms that Coulomb cooling in the CM layer can indeed dominate, at least for electrons with high pitch angle, which mirror close to the loop apex.

A more complete analysis (Section \ref{sec:loss_rate_comparison}) of the relative energy losses when integrated along the loop shows, however, that more field-aligned electrons that mirror in the stronger magnetic field close to loop footpoints have gyro-emission that competes or even exceeds collisional cooling from CM crossings
(see Equation \ref{eq:rabR1}).
Numerical computations in Section \ref{sec:numerical_comp} then give results for full models with an initially gyrotropic distribution introduced over a range of loop heights, confirming that residual gyro-emission is generally concentrated around the loop footpoints in the magnetic polar regions near the star.
Section \ref{sec:discussion} discusses the potential application for radio emission from cooler objects such as brown dwarfs and also the need for resolved radio imaging of magnetic hot stars.
We conclude (Section \ref{sec:summary}) with a brief summary and outline for future work.


\section{Cooling by collisions with thermal electrons}

\subsection{General scalings}\label{subsec:general_scaling}

In addition to the gyro-cooling examined in Paper\,I, non-thermal electrons can also cool by the energy exchange from Coulomb collision with an ambient population of thermal electrons of much lower energy
\citep[see][e.g., their section 8.4.3]{2009LNP...778..269G}.
For potentially relativistic electrons with speed $v= \beta c$  near the speed of light $c$, {\color{black}thus with} Lorentz factor $\gamma \equiv 1/\sqrt{1-\beta^2}$ and kinetic energy E=$(\gamma -1)m_\mathrm{e}c^2$ {\color{black}($m_\mathrm{e}$ is the mass of an electron)}, the
cooling time for collisions with thermal electrons of number density $n_\mathrm{e}$ is 
\citep[][see their {\color{black}Equation} 1]{1981ApJ...251..781L}
\beq
t_\mathrm{c} 
\equiv \frac{E}{dE/dt}
= \frac{\beta (\gamma-1)}{n_\mathrm{e} 4\pi r_\mathrm{o}^2 \ln \Lambda \, c }
= 1.7 \times 10^{12} \, {\rm s} \, \frac{\beta (\gamma-1)}{ \, n_\mathrm{e}\ {\rm cm}^3} \, ,
\label{eq:tcdef}
\eeq
where for electron charge $q_\mathrm{e}$ and mass $m_\mathrm{e}$, $r_\mathrm{o} = q_\mathrm{e}^2/m_\mathrm{e} c^2$ is the classical electron radius, and the latter evaluation assumes a characteristic value $\ln \Lambda \approx 20$ for the Coulomb logarithm \citep[which accounts for the cumulative effect of many small-angle scatterings, {\color{black}$\Lambda^{-1}$ represents the minimum angle of deflection in the Coulomb integral,}][]{1981ApJ...251..781L}.

Assuming an isotropic pitch-angle distribution with \\
$\left < \sin^2 \alpha \right > = 2/3$\footnote{{\color{black}The averaging is performed over the solid angle $d\Omega=\sin\theta d\theta d\phi$. In this context, $\theta\equiv \alpha$, where $\alpha$ is the pitch angle.}} {\color{black}(where $\alpha$ is the pitch-angle)}, comparison with {\color{black}Equation 2} of Paper\,I shows that the ratio of the Coulomb to gyro-synchrotron cooling times {\color{black}$t_\mathrm{e}$} is given by
\beq
\frac{t_\mathrm{c}}{t_\mathrm{e}} \approx 2.2 \times
10^3 \, \frac{(\gamma^2 -1) \beta B^2}{ n_\mathrm{e}} \, ,
\label{eq:tcbte}
\eeq
where $B$ {\color{black}is the magnetic field strength. Both $B$ and $n_\mathrm{e}$ are} taken to be in CGS units.

\begin{figure}
\includegraphics[width=0.49\textwidth]{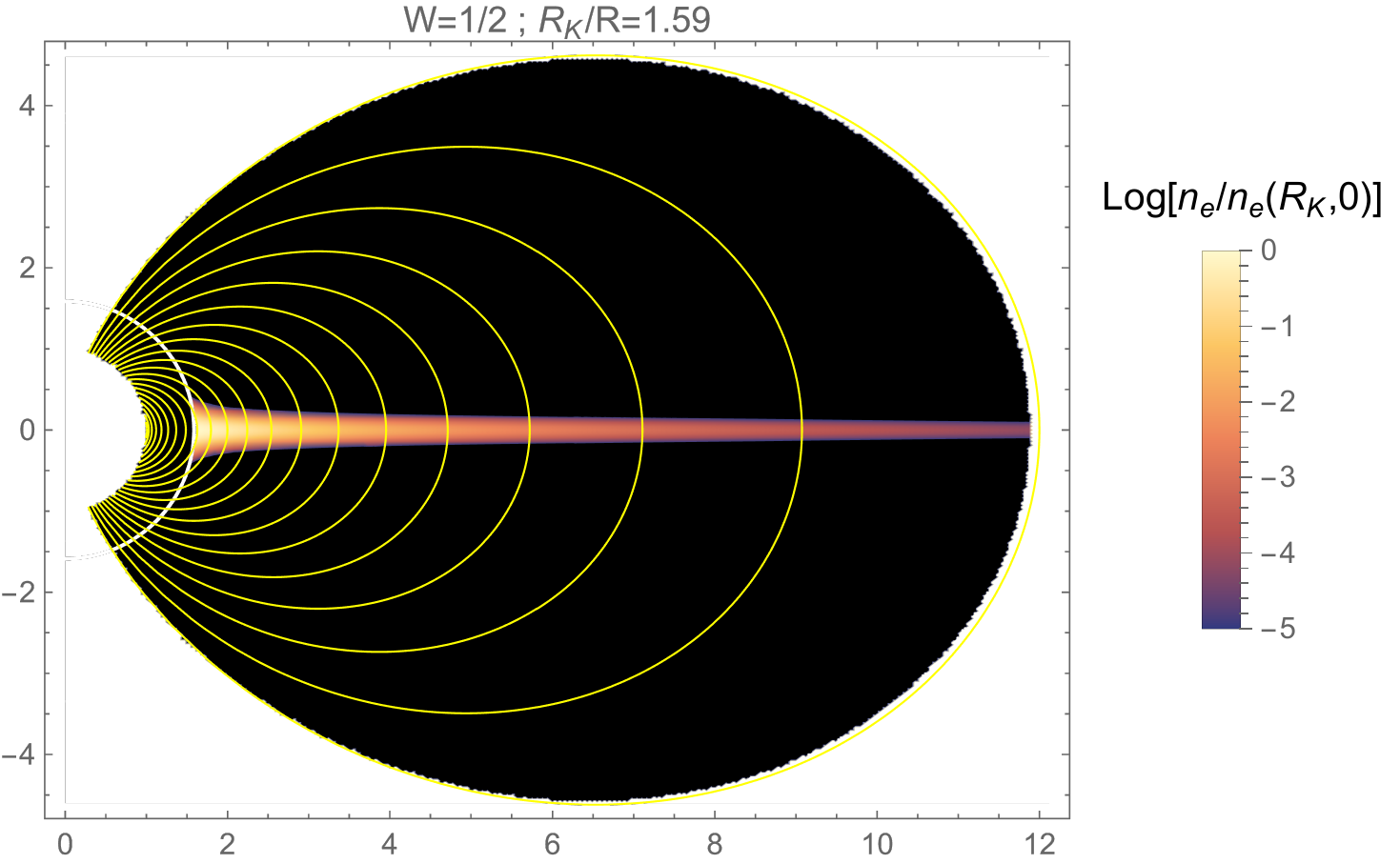}
\caption{
{\color{black}
For our assumed model of a centrifugal magnetosphere (CM) limited by centrifugal breakout (CBO), the color plot shows the log of electron density normalized to its maximum value in the equator at the Kepler radius, which for this model with critical rotation fraction $W=1/2$, occurs at $R_K/R=W^{-2/3}= 1.59$ (denoted here by the white circle). The yellow contours show magnetic field lines extending to an outer radius $r=12 R$, {\color{black} with spacing set to follow the  field strength.}}}
\label{fig:CBOCM}
\end{figure}

\subsection{Cooling in the dense CM layer}\label{subsec:cooling_CM}

To proceed, we need to specify a model for the density distribution of thermal electrons.
For CMs of hot stars, the strongest collisional cooling in the extended CM should be in the dense, compressed layer near the tops of magnetic loops.
For the simple axisymmetric case of a rotationally aligned dipole field, the {\color{black}CBO} analysis by \citet[][see their Equation\ 7]{owocki2020} provides a scaling for the radial variation of mass column density in this CM layer.
Through the mean mass per electron $\mu_\mathrm{e} = 1.16 m_\mathrm{p}$, {\color{black}where $m_\mathrm{p}$ is the mass of a proton}, this can be used to derive an associated electron column density of this thin, dense CM layer,
\beqa
N_\mathrm{e} &\approx & 0.3 
\frac{B_{\rm K}^2}{4 \pi \mu_\mathrm{e} g_{{\rm K}}} \, 
\left ( \frac{r_\mathrm{a}}{R_{\rm K}} \right )^{-p} 
\label{eq:NeCBO0}
\\
&\approx& 4.85 \times 10^{22} {\rm cm}^{-2} \, \frac{B_\mathrm{kG}^2}{g_4} 
\left ( \frac{R_{\rm K}}{R} \right )^{-4}
\left ( \frac{r_\mathrm{a}}{R_{\rm K}} \right )^{-5} 
\, ,
\label{eq:NeCBO}
\eeqa
where $B_{\rm K} $and $g_{\rm K}$ are the magnetic field strength and stellar gravity at the Kepler radius {\color{black}$R_\mathrm{K}$}, and the power index $p$ allows for generalizations from the original $p=6$ assumed in {\color{black}Equation}\ 7 of \citet{owocki2020}; {\color{black}$r_\mathrm{a}$ is the apex radius of the magnetic field loop under consideration.}

Following \citet{Berry2022} and \citet{ud-doula2023}, the latter equality
instead takes $p=5$, and provides numerical scalings in terms of the scaled surface gravity $g_4 \equiv g_\ast/(10^4 \,$cm/s$^2$), and the polar surface field in kG, $B_\mathrm{kG}\equiv B_\mathrm{p\ast
 }/$kG.
 For a star with critical rotation fraction $W$ at its equatorial surface, $R_{\rm K}/R = W^{-2/3}$.

For our assumed standard model with $W=1/2$ (and thus $R_K=1.59 \, R$), Figure \ref{fig:CBOCM} illustrates the spatial variation of electron density to an outer radius $r=12 ~ R$, with yellow curves showing dipole field lines.

 For high-energy electrons with speed $v$ along a path $s$ through this CM layer at the loop apex, application of {\color{black}Equation} (\ref{eq:tcdef}) implies that the fractional loss of energy is set by the collision depth,
 \beqa
 \tau_\mathrm{c} = \int \frac{ds}{v \, t_\mathrm{c}} &=& \frac{N_\mathrm{e} 4 \pi r_\mathrm{o}^2 \ln \Lambda }{\beta^2 (\gamma -1)} 
 \label{eq:tauc1}
 \\
 &\approx& 0.96 \, \frac{B_\mathrm{kG}^2}{g_4 \beta^2 (\gamma-1)} (2W)^{8/3} \left ( \frac{r_\mathrm{a}}{R_{\rm K}} \right )^{-5} 
\, .
 \label{eq:tauc2}
\eeqa
For each pass through the CM layer, the fractional energy loss is $1-e^{-\tau_\mathrm{c}}$.

\subsection{Reduction of gyro-emission}\label{subsec:reduced_gyrocool}
Let us next consider how such cooling from Coulomb collision competes with the gyro-cooling over a given mirror cycle. 
Equation 11 of Paper\,I defined a constant $k$ to
characterise the energy loss to gyro-cooling during each mirror cycle.
Noting that the Thomson cross section $\sigma_\mathrm{T} = (8/3) \pi r_\mathrm{o}^2$ and casting this in terms of these Kepler-radius parameter scalings, we have 
\beqa
~~~~~~~~~~k &=&  
\frac{4 B_{\rm K}^2 R_{\rm K} r_\mathrm{o}^2}{3 m_\mathrm{e} c^2 \beta_0} \, 
\left ( \frac{r_\mathrm{a}}{R_{\rm K}} \right )^{-5} 
\label{eq:kK0}
\\
&=&  0.0032 \, \frac{ B_\mathrm{kG}^2 R_{12} }{\beta_0} (2W)^{10/3}
\left ( \frac{r_\mathrm{a}}{R_\mathrm{K}} \right )^{-5} 
\, ,
\label{eq:kK}
\eeqa
where $R_{12} \equiv R/(10^{12}$cm) and $\beta_0 \equiv v_0/c$.

Comparison of Equation\ (\ref{eq:kK}) with (\ref{eq:tauc1}) shows that, quite remarkably,  both gyro-cooling along magnetic loops and Coulomb cooling in the CM layer have the {\it same} dependence on the magnitude and radial variation of magnetic field, implying then that their ratio is a {\it global} parameter that is  fully {\it independent} of the field,
\beq
\boxed{
\frac{k}{\tau_\mathrm{c}} 
\approx 
C_\mathrm{k}
\, (\beta^2/\beta_0) (\gamma -1) \, 
(2W)^{2/3}\, 
{\mathcal M}/{\mathcal R}
}
\, 
\label{eq:kbtauc}
\eeq
where ${\mathcal M}/{\mathcal R}$ is
the star's ratio of mass to radius in solar units.
Combining (\ref{eq:NeCBO0}), (\ref{eq:tauc1}) and (\ref{eq:kK0}), we find the coefficient $C_{k}$ has a very small value, set by the scaling
\beq
C_\mathrm{k} \equiv
\frac{2.8}{\ln \Lambda} \,\frac{G M_\odot}{R_\odot c^2} \, \frac{\mu_\mathrm{e}}{m_\mathrm{e}}
\approx 6.3\times 10^{-4}
\, ,
\label{eq:Ckdef}
\eeq
where the latter evaluation again uses $\ln \Lambda = 20$ and $\mu_\mathrm{e} = 1.16 \, m_\mathrm{p}$.

For typical scaled parameters of order unity, equations (\ref{eq:kbtauc} and (\ref{eq:Ckdef}) indicate that collisional cooling should dominate over
gyro-cooling, at least for electrons with large pitch angle that mirror near the loop apex.

However, for lower pitch angles that mirror closer to the loop footpoints, where the much stronger field implies
a much stronger gyro-emission, quantifying the net reduction requires  integration over the full loop, 
as discussed next.

\section{Analytic comparison of loss rates}\label{sec:loss_rate_comparison}

Building upon the Paper\,I analysis of pure gyro-cooling of energetic electrons introduced near the loop tops of closed dipole field lines, let us now add the competing energy loss from collisional cooling within a dense CM layer near the loop apex.
For potentially relativistic electrons, the relevant time evolution for energy $e$, magnetic moment $p$ and co-latitude cosine $\mu$ are given by equations (A5), (A6), and (A3) of Paper\,I, with the energy loss now supplemented by the collision cooling rate that we now derive.

\subsection{Energy loss rate from Coulomb cooling}\label{subsec:energy_loss_rate}

Using {\color{black}Equation}\ (\ref{eq:tcdef}), the rate {\color{black}of} energy loss from Coulomb collisions is
\beq
\frac{dE}{dt} = - \frac{E}{t_\mathrm{c}} = - \frac{m_\mathrm{e} c^2 n_\mathrm{e} c 4 \pi r_\mathrm{o}^2 \ln \Lambda  }{\beta }
\, .
\eeq
Using $E=(\gamma-1)m_\mathrm{e} c^2$ and $e=(\gamma-1)/(\gamma_0-1)$, for initial Lorentz factor $\gamma_0$, we have
\beq
\frac{de}{dt} = - \frac{n_\mathrm{e} c 4 \pi r_\mathrm{o}^2 \ln \Lambda }{\beta (\gamma_0-1)}
\eeq
Scaling the time in units of the initial apex crossing time $t_\mathrm{a}=r_\mathrm{a}/v_0$, we find
\beqa
{\dot e}_\mathrm{c} \equiv \frac{de}{dt} t_\mathrm{a} &=& - \frac{n_\mathrm{e} c 4 \pi r_\mathrm{o}^2 \ln \Lambda}{\beta (\gamma_0-1)}
\frac{r_\mathrm{a}}{v_0}
\\
&=& - \frac{\tau_\mathrm{c} \beta (\gamma-1)}{\beta_0 (\gamma_0-1)} \, \frac{{\rm e}^{-(\mu/\mu_\mathrm{H})^2}}{\sqrt{\pi} \mu_\mathrm{H}}
\, ,
\eeqa
where the second expression uses (\ref{eq:tauc1}) and invokes the gaussian density stratification of the CM layer over a scale height
given by \citep[see][their {\color{black}Equation}\ 4 ]{Owocki2018}
\beq
H = \frac{\sqrt{2} c_\mathrm{s}/\Omega}{\sqrt{3-2(R_\mathrm{K}/r_\mathrm{a})^3}} 
\, ,
\eeq
with $\Omega$ the stellar rotation frequency.
For $r_\mathrm{a} \ge R_\mathrm{K}$, the corresponding scale in co-latitude $\mu$ is
\beq
\mu_\mathrm{H} \equiv  \frac{H}{r_\mathrm{a}} = \frac{0.063}{ \sqrt{{\mathcal M}/{\mathcal R}}\sqrt{3W^2-2(R/r_\mathrm{a})^3}}
\, \frac{R}{r_\mathrm{a}}
\, ,
\label{eq:muHdef}
\eeq
where the latter evaluation assumes an isothermal sound speed $c_\mathrm{s} = 20$\,km/s.
For standard parameters $2W = {\mathcal M}/{\mathcal R} = 1$, we find $\mu_\mathrm{H} = 0.026 $ for $r_\mathrm{a}=3 R$ and  $\mu_\mathrm{H}=0.012 $ for $r_\mathrm{a}=6 R$.

Finally,  using (\ref{eq:kbtauc}) to {\color{black}eliminate} $\tau_\mathrm{c}$ in favor of the gyro-cooling parameter $k$, we find
\beq
\boxed{
{\dot e_\mathrm{c}}
= -\frac{1600 \, k}{\beta (\gamma_0-1) (2W)^{2/3} {\mathcal M}/{\mathcal R} }\, \frac{{\rm e}^{-(\mu/\mu_\mathrm{H})^2}}{\sqrt{\pi} \mu_\mathrm{H}}
}
\label{eq:edotc}
\, .
\eeq 

\subsection{Quantitive scaling of collisional cooling vs.\ gyro-cooling}

By comparison, {\color{black}Equation}  (A5)  of Paper\,I gives the relativistic form for the competing gyro-cooling emission term. Noting that the relativistic correction factor in curly brackets just becomes $(\gamma +1)/2$, 
their ratio can be characterized by
\beq
\frac{\left < {\dot e}_\mathrm{e} \right >}{\left < {\dot e}_\mathrm{c}\right >} = 3.1 \times 10^{-4}  \beta  (\gamma_0-1)(\gamma+1)\left <pb^3 \right >
\, (2W)^{2/3} {\mathcal M}/{\mathcal R} 
\,  ,
\label{eq:edotebedotc}
\eeq
where the angle brackets denote the cumulative time average (see {\color{black}Equation}\ \ref{eq:pb3int} below) from the apex $\mu=0$ to the mirror point $\mu=\mu_\mathrm{m}$.

In the common simple case that both energy losses are small across a single mirror loop crossing, implying then that
$\gamma \approx \gamma_0$ and $\beta \approx \beta_0$, we can approximate the ratio of gyro-emission to Coulomb loss as
\beq
\boxed{
\frac{\delta e_\mathrm{g}}{\delta e_\mathrm{c}} \approx 
3.1 \times 10^{-4} {\beta_0^3}{\gamma_0^2} \, \left <pb^3 \right >
\, (2W)^{2/3} {\mathcal M}/{\mathcal R}
}
 \, .
\label{eq:degbdec}
\eeq
Note that for non-relativistic electrons with $\beta_0 < 1$, this ratio {\it declines} as $\delta e_\mathrm{g}/\delta e_\mathrm{c} \sim  \beta_0^3$,
 implying strong dominance of Coulomb cooling.

On the other hand, in the relativistic regime $\gamma_0 > 1$ , the ratio {\it increases} as $\delta e_\mathrm{g}/\delta e_\mathrm{c} \sim \gamma_0^2$,
implying a much reduced importance of Coulomb cooling for such relativistic electrons.

For electrons with large pitch angles, which mirror near the loop top, $\left <pb^3 \right >$ remains of order unity, showing again that, unless the electrons 
are highly relativistic $\gamma_0 \gg 1$, losses by Coulomb collisions should dominate over gyro-cooling.

But this factor $\left < pb^3 \right >$ can become large for field-aligned pitch angles, for which the mirror 
radius $r_\mathrm{m} \ll r_\mathrm{a}$, implying an increase field strength that scales as $b \sim 1/r_\mathrm{m}^3$, allowing a potentially
significant gyro-cooling even for the mildly relativistic case, $\gamma_0 \gtrsim 1$.

\subsection{Pitch-angle dependence}\label{subsec:alpha_dep_analytical}

To quantify this pitch-angle dependence, let us again assume we can neglect the energy losses for a single mirror cycle.
By conservation of the scaled magnetic moment $p = \sin^2 \alpha_\mathrm{a}$, the apex-scaled field strength at the mirror point then just varies as
\beq
b_\mathrm{m} \equiv \frac{B_\mathrm{m} }{B_\mathrm{a}} = \frac{1}{\sin^2 \alpha_\mathrm{a}} = \frac{1}{p} 
\, .
\label{eq:bdef}
\eeq
Since an electron mirroring from the loop apex spends the greatest time near its mirror radius, one might initially estimate  that
$
\left < p b^3 \right > 
\approx b_\mathrm{m}^2 \approx (\sin \alpha_\mathrm{a} )^{-4}
\, .
$ 

\begin{figure}
\includegraphics[width=0.49\textwidth]{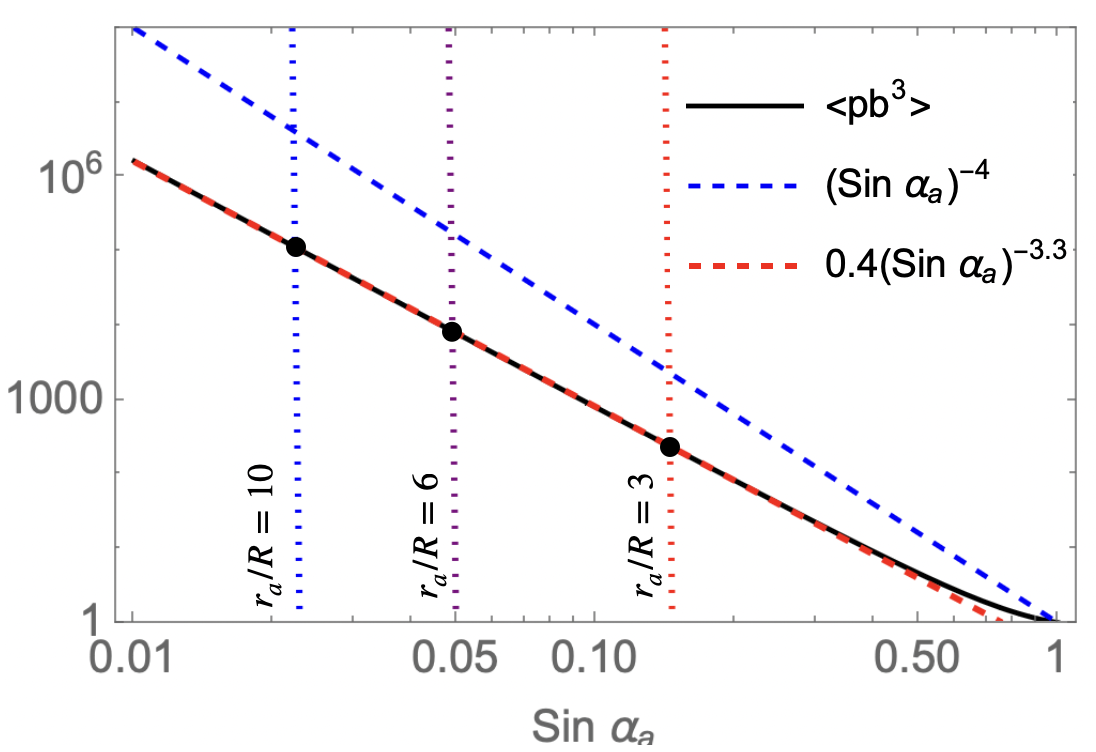}
\includegraphics[width=0.49\textwidth]{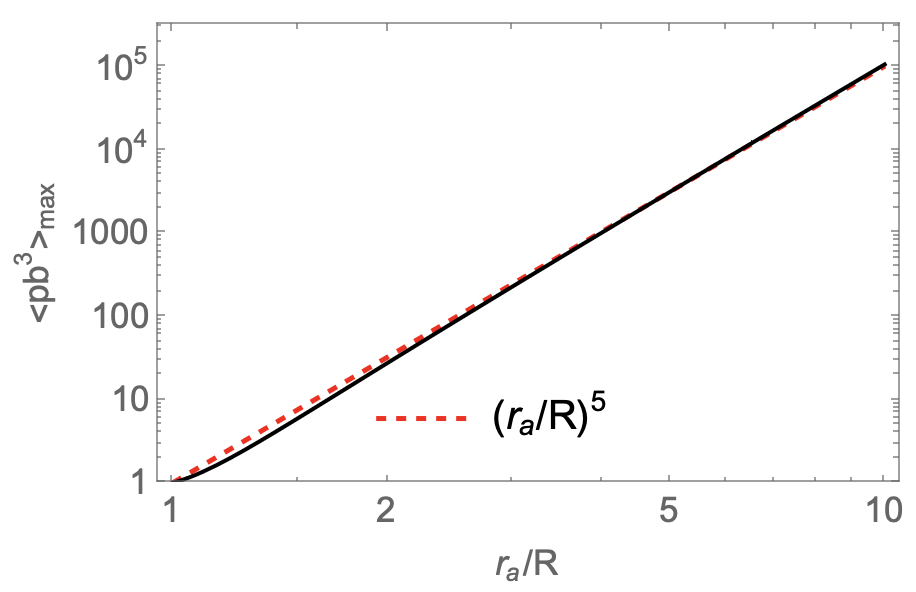}
\caption{Top: On a log-log scale, the black curve plots the mirror-cycle time-average of $\left < p b^3 \right >$ computed from (\ref{eq:pb3int} vs. sine of apex pitch angle $\sin \alpha_\mathrm{a}$. The red and blue dashed lines show that $(\sin \alpha)^{-3.3}$ is a much better fit to $\left < p b^3 \right >$ than the simple estimate $(\sin \alpha)^{-4}$. The vertical dotted lines mark the minimum apex pitch angle for the mirror radius $r_\mathrm{m}$ to remain above the stellar radius $R$ for the labeled values of apex radius $r_\mathrm{a}$. The black dots thus mark the maximum possible value of $\left < p b^3 \right >$ for loops with these apex radii.
Bottom: The black curve now shows this $\left < p b^3 \right >_\mathrm{max}$ plotted vs.\ $r_\mathrm{a}/R$. The red dashed curve shows that this maximum is quite well fit by $(r_\mathrm{a}/R)^5$.
}
\label{fig:pb3vssina}
\end{figure}

But using $p = \sin^2 \alpha_\mathrm{a}$, the full average over the mirror time $t_\mathrm{m}$ can be readily computed from numerical integration over a mirror cycle,
\beq
\left < p b^3 \right > = \frac{\sin^2 \alpha_\mathrm{a}}{t_\mathrm{m}}  \int_0^{t_\mathrm{m}} b(t)^3 \, dt \, .
\label{eq:pb3int}
\eeq
In the upper panel of Figure \ref{fig:pb3vssina}, the black curve  plots the resulting $\left < p b^3 \right >$ vs. $\sin \alpha_\mathrm{a}$ on a log-log scale. Comparison with the red and blue dashed lines show that, instead of the above simple estimate $(\sin \alpha)^{-4}$, a much better fit to the full integral (\ref{eq:pb3int}) is given by 
\beq
\left < p b^3 \right > \approx 0.4 
(\sin \alpha)^{-3.3}
\, .
\label{eq:pb3approx}
\eeq

The vertical dotted lines in Figure \ref{fig:pb3vssina} mark the minimum apex pitch angle for the mirror radius $r_\mathrm{m}$ to remain above the stellar radius $R$ for the labeled values of apex radius $r_\mathrm{a}$. The black dots thus mark the maximum possible value of $\left < p b^3 \right >$ for loops with these apex radii.

The black curve in the lower panel of Figure \ref{fig:pb3vssina} then plots this maximum vs.\ $r_\mathrm{a}/R$. Comparison with the red dashed line shows that, quite remarkably,  the maximum has the approximate simple scaling\footnote{If we approximate the mirror radius ratio $r_\mathrm{m}/r_\mathrm{a} \approx (\sin \alpha_\mathrm{a})^{-2/3}$ and set $r_\mathrm{m}=R$ for the minimum pitch angle, application in (\ref{eq:pb3approx}) gives this scaling in (\ref{eq:pb3maxra5}).}
\beq
\left < p b^3 \right >_\mathrm{max} \approx \left ( \frac{r_\mathrm{a}}{R} \right )^5
\, .
\label{eq:pb3maxra5}
\eeq
This indicates that in high-lying loops with large ratio of apex to stellar radius $r_\mathrm{a}/R$, electrons with low pitch angle $\alpha_\mathrm{a}$ that mirror just above the stellar radius can have $\left <pb^3 \right >$ that approach the large maximum set by {\color{black}Equation} (\ref{eq:pb3maxra5});
this can thus compensate for the small numerical factor in {\color{black}Equation} (\ref{eq:degbdec}).

For example, applying (\ref{eq:pb3maxra5}) into (\ref{eq:degbdec}), we can solve for the apex radius ratio that would give $\delta e_\mathrm{g}/\delta e_\mathrm{c} =1$,
\beq
\boxed{
\left ( \frac{r_\mathrm{a}}{R} \right )_1 \approx
5 \left ( 
{\beta_0^3}{\gamma_0^2} \, 
(2W)^{2/3} {\mathcal M}/{\mathcal R}
\right )^{-1/5}
}
\, .
\label{eq:rabR1}
\eeq
The upshot here is that for even mildly relativistic electrons ($\beta_0 \approx 0.76$, $\gamma_0 \approx 1.5$) for which $\beta_0^3 \gamma_0^2 \approx 1$,
loops with an apex radius near $r_\mathrm{a} \approx 5 R$ can have gyro-cooling that competes with collisional cooling, despite the small numerical value of the factor ($C_\mathrm{k}/2$) in (\ref{eq:degbdec}).

\section{Numerical computations}\label{sec:numerical_comp}
\begin{figure*}
\includegraphics[width=0.49\textwidth]{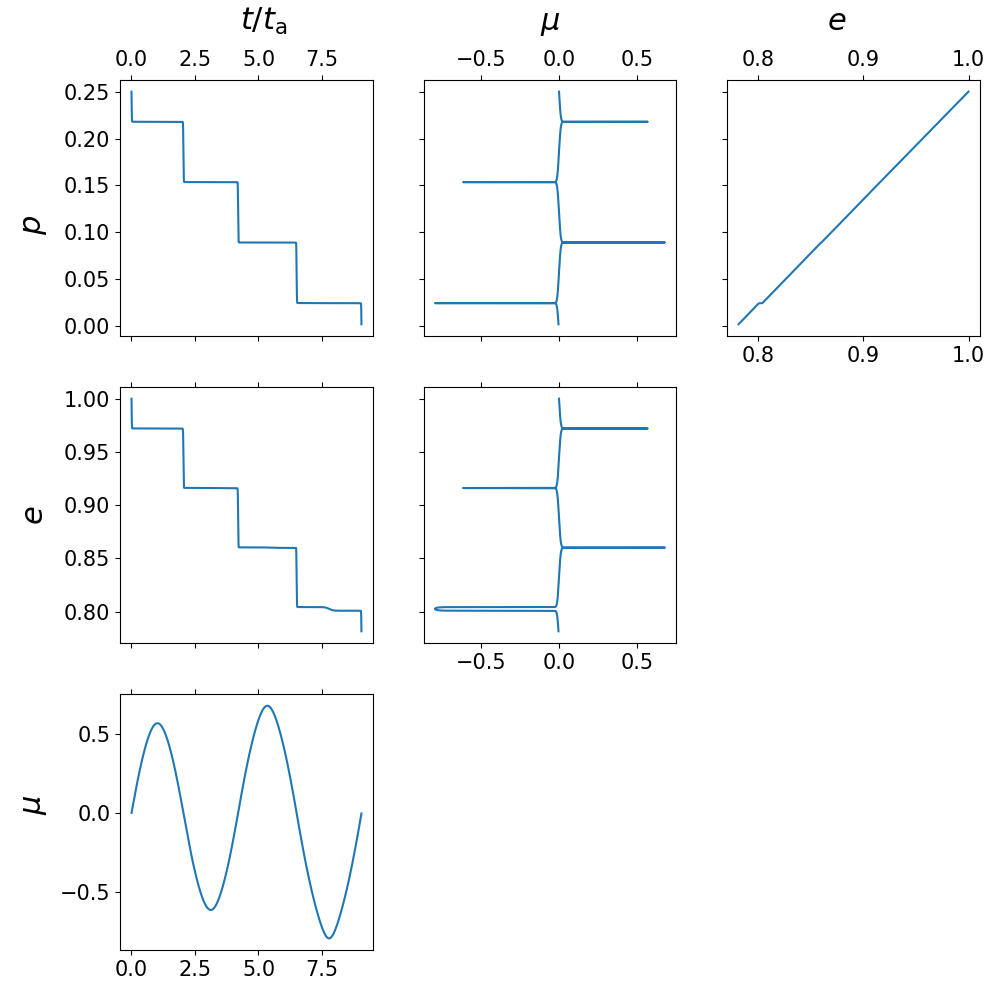}
\includegraphics[width=0.49\textwidth]{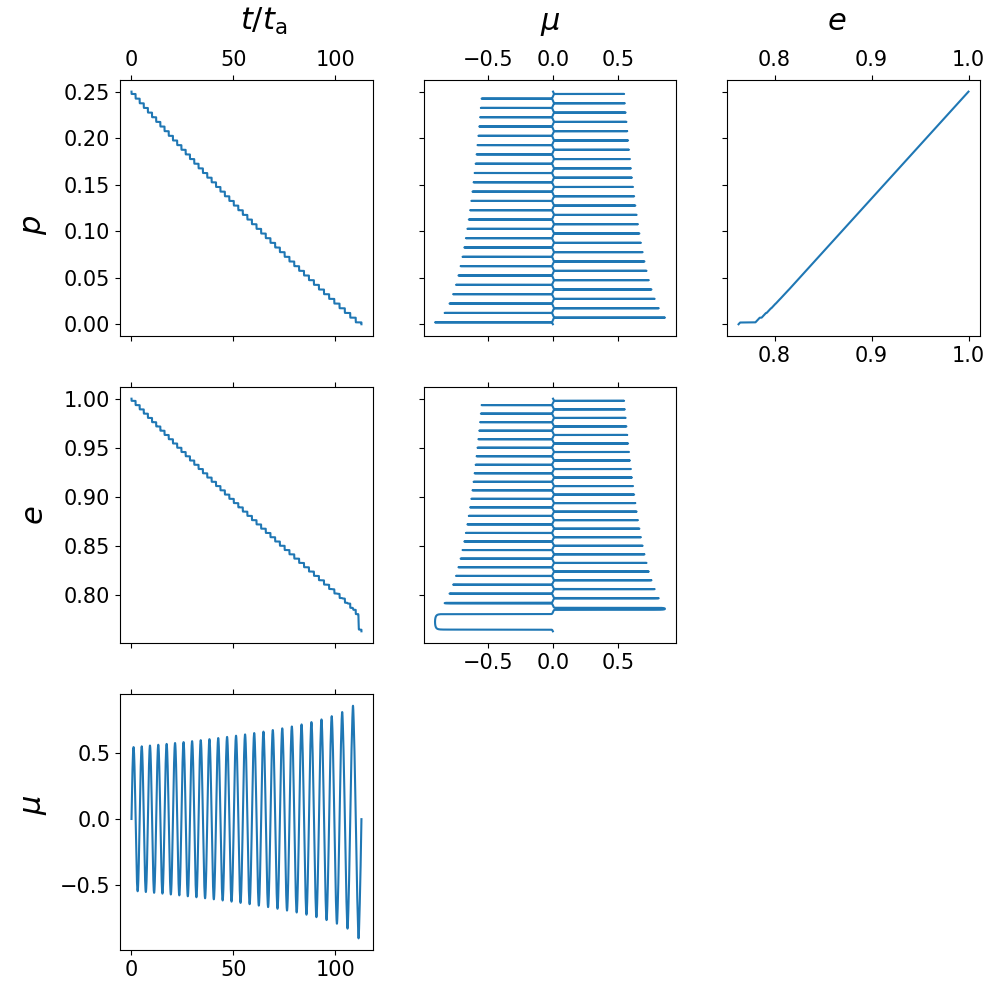}
\caption{
For a mildly relativistic electron with initial Lorentz factor $\gamma_0 = 1.5$ and initial pitch angle $\alpha_\mathrm{a} = 30^o$, the variation of magnetic moment $p$, energy $e$, and latitudinal cosine $\mu$ plotted vs. {\color{black}the dimensionless time $t/t_{\rm a} = v_0 t/r_{\rm a}$} (left columns),
latitude $\mu$ (middle columns), and magnetic moment $p$ (right columns), for apex radii $r_\mathrm{a}=6R$ (left) and $r_\mathrm{a}=10R$ (right).  The weaker field strength and gyro-cooling in the right set leads to many more mirror cycles on the right vs. left. But the net relative importance of collisional  vs. gyro-synchrotron cooling across the CM layer -- shown by the drops in $e$ and $p$ across $\mu=0$ -- are the same in the left vs. right cases.
}
\label{fig:fig3}
\end{figure*}

Building upon the above  analytic analyses, let us now extend the numerical computations of Paper\,I to include the effects of Coulomb collisional cooling, as given by {\color{black}Equation} (\ref{eq:edotc}).
All models presented assume a standard stellar parameter set with $2W = {\mathcal M}/{\mathcal R} = 1$.

\subsection{Single electron evolution}

In analogy with Figure 1 of Paper\,I, Figure \ref{fig:fig3} here illustrates the evolution of the energy $e$, magnetic moment $p$, and latitudinal cosine $\mu$ for an electron introduced at loop apex radius $r_\mathrm{a}$ with initial pitch angle $\alpha_\mathrm{a} = 30^o$, and with an initially relativistic Lorentz factor $\gamma_\mathrm{0}=1.5$.
The left and right panels contrast results for $r_\mathrm{a}=6 R$ vs.\ $r_\mathrm{a}=10R$.
For the higher apex radius, the much lower value of the magnetic field, and thus of the gyro-cooling constant $k$,
leads to many more mirror cycles than for the lower apex radius $r_\mathrm{a}= 6R$.

One key difference from the case without collisional cooling is that here the time evolution of the particle is governed by the evolution of $p$, since once $p$ becomes zero (i.e. the particle's velocity becomes parallel to the magnetic field), it will no longer lose energy by radiation. Thus, any further loss of energy occurs only by collisions, which do not produce observable radio emission. For the initial pitch angle of $30^\circ$, collisional cooling always dominates over gyrocooling, which is consistent with the analytical analysis in \S\ref{subsec:alpha_dep_analytical} with $\langle pb^3\rangle \approx 4$ following Eq. \ref{eq:pb3approx}.

\begin{figure}
    \centering
    \includegraphics[width=0.49\textwidth]{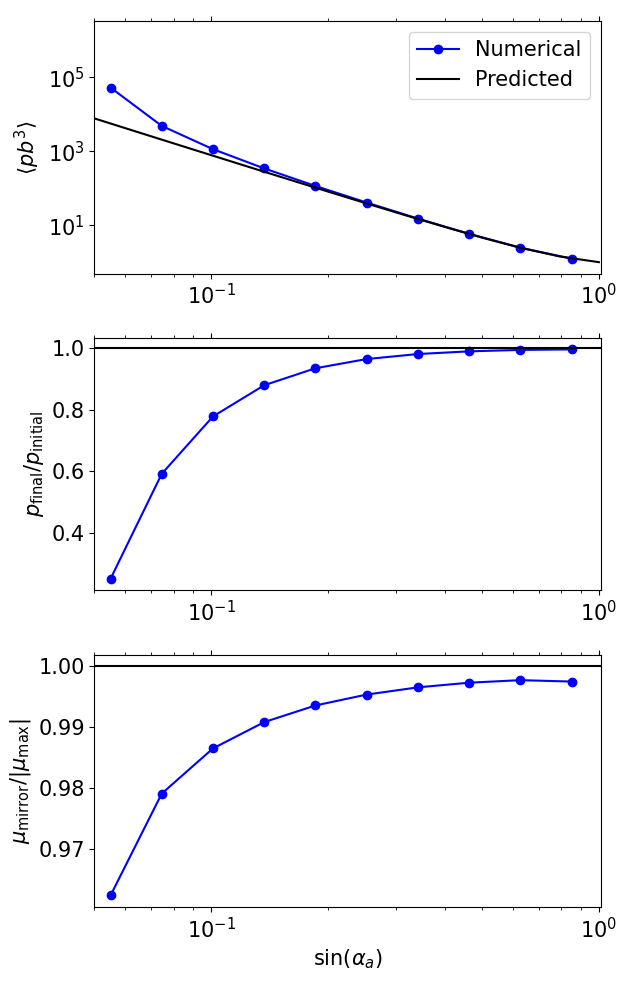}
    \caption{\textit{Top:} Comparison between numerically computed values of $\langle pb^3\rangle$ (as defined  in \S\ref{subsec:alpha_dep_analytical}) with the 
    analytically predicted ones, plotted as a function of initial pitch angle for a fixed apex radius of $10\,R$. \textit{Middle:} The ratio between final value of $p$ to its initial values over the time ranges considered for the top panel. \textit{Bottom:} Ratio between predicted mirroring cosine to the `actual' (numerical) mirroring cosine.}
    \label{fig:pb3_numerical}
\end{figure}

For the case with $r_\mathrm{a}=10\,R$, the top panel of Figure \ref{fig:pb3_numerical} compares the numerical values $\langle pb^3 \rangle$ with the analytic predictions from Eq. \ref{eq:pb3approx}.
For the relatively large values of initial pitch angle, the two agree remarkably well; but for the smaller values of initial pitch angles, the numerically obtained values are higher than predicted. The reasons for this are illustrated in the middle panel, which shows that the final to initial value of $p$ decreases sharply with lower initial pitch angle $\sin \alpha_\mathrm{a}$, demonstrating then that
the analytic assumption that $p$ does not change significantly over a mirror cycle is simply {\em not} valid for such low initial pitch angles. 
The bottom panel 
shows that this enhances electron penetration into the lower magnetosphere, thus increasing values of $\langle p b^3 \rangle$ from what's predicted from the analytic analysis.

Another important difference between the analytical and numerical results is that, while the former shows the minimum value of $\sin\alpha_\mathrm{a}$ for $r_\mathrm{a}=10\,R$ to be $\approx 0.024$, in the numerical simulation, electrons with such low pitch angle completely lose their magnetic moment $p$ by the time they come out of the dense plasma at the apex, and so no longer contribute to radiation. 

A first order estimate of the minimum $\sin\alpha_\mathrm{a}$ for a given value of $k$ and $r_\mathrm{a}$ can be obtained as follows. For electrons close to the apex radius, gyrocooling is negligible compared to collisional cooling. Thus, we can write:
\begin{align*}
    \frac{dp}{dt}&\approx \frac{1}{b}\left(\frac{de}{dt}\right)_\mathrm{col}
\end{align*}
Changing the variable from $t$ to $\mu$, we get:
\begin{align*}
    \frac{dp}{d\mu}&\approx -\frac{1}{b}\frac{\sqrt{1+3\mu^2}}{\sqrt{e-pb}}\frac{1600 k}{\beta(\gamma_0-1)}\frac{\exp{\{-(\mu/\mu_\mathrm{H})^2\}}}{\sqrt{\pi}\mu_\mathrm{H}}\\
    &\approx -\frac{1600 k}{\sqrt{e-p}\beta(\gamma_0-1)\sqrt{\pi}\mu_\mathrm{H}}\exp{\{-(\mu/\mu_\mathrm{H})^2\}}
\end{align*}
where we have set $b=1$ and $\sqrt{1+3\mu^2}\approx 1$ for regions close to the apex. Also, under the assumptions:
\begin{align}
    \dot{p}&=\dot{e}\\
    \Rightarrow p&=p_0+\dot{e}dt\approx p_0+e-1\\
    \Rightarrow e-p& \approx 1-p_0
\end{align}
where $p_0$ is the initial value of $p$. Thus, we can write,
\begin{align*}
    p &= p_0-\frac{1600 k}{(\gamma_0-1)\sqrt{1-p_0}}\int_0^{\mu}\frac{\exp{\{-(\mu/\mu_\mathrm{H})^2\}}}{\beta\sqrt{\pi}\mu_\mathrm{H}} d\mu
\end{align*}
Here $\mu \gg \mu_\mathrm{H}$ represents a value just outside the dense plasma layer. Now $p$ will be greater than zero, if 

\begin{align*}
    p_0&>\frac{1600 k}{(\gamma_0-1)\sqrt{1-p_0}}\int_0^{\mu}\frac{\exp{\{-(\mu/\mu_\mathrm{H})^2\}}}{\beta\sqrt{\pi}\mu_\mathrm{H}} d\mu \\
    &\geq \frac{1600 k}{\beta_0(\gamma_0-1)\sqrt{1-p_0}}\int_0^{\mu}\frac{\exp{\{-(\mu/\mu_\mathrm{H})^2\}}}{\sqrt{\pi}\mu_\mathrm{H}} d\mu 
\end{align*}
For $\mu \gg \mu_\mathrm{H}$, the integral equals $1/2$. Also, since we are considering the cases where the initial $p_0$ is already small, we can set $1-p_0\approx 1$. Thus, we get:
\begin{align}
    p_0&>\frac{800 k}{\beta_0(\gamma_0-1)} \nonumber \\
    |\sin\alpha_\mathrm{a}|&> {\left( \frac{800 k}{\beta_0(\gamma_0-1)} \right)}^{1/2}
\end{align}
For our numerical simulation, we have used $B_*=3.8$ kG, $R=2\, R_\odot$ and $W=0.5$ (parameters similar to CU\,Vir), which gives $|\sin\alpha_\mathrm{a}|> 0.043$ for $r_\mathrm{a}=10\, R$. The minimum value of $\sin\alpha_\mathrm{a}$ in Figure \ref{fig:pb3_numerical} is 0.055.

\begin{figure}
    \centering
    \includegraphics[width=0.49\textwidth]{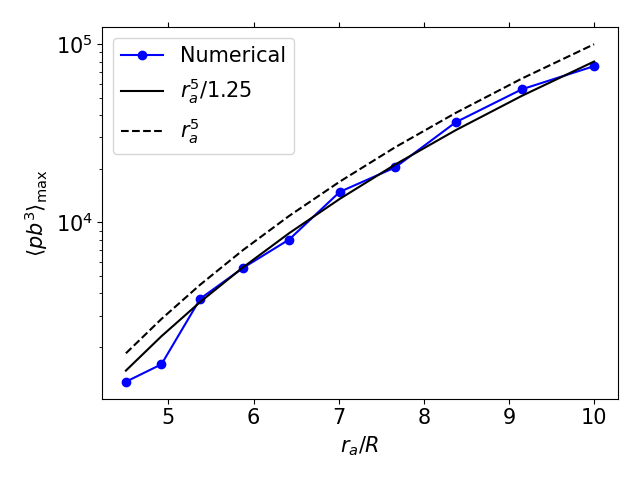}
    \caption{Numerically computed $\langle pb^3\rangle_\mathrm{max}$ as a function of $r_\mathrm{a}/R$.}
    \label{fig:pb3_max_vs_ra}
\end{figure}

\begin{figure}
    \centering
    \includegraphics[width=0.49\textwidth]{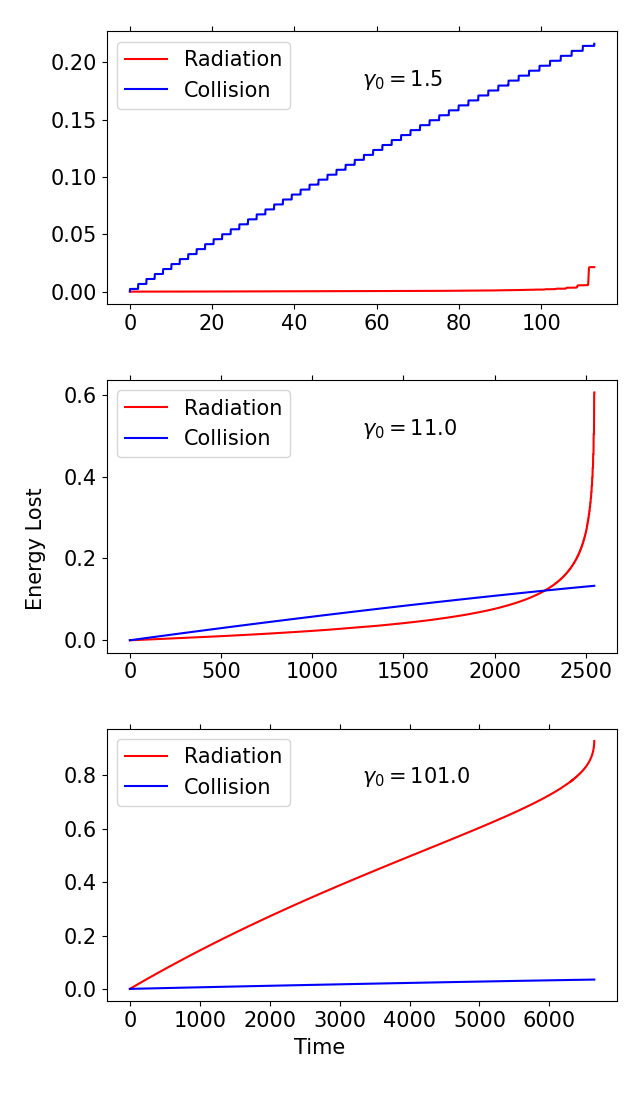}
    \caption{Time evolution of energy lost by radiation and collision for an electron with initial pitch angle of $30^\circ$, injected at the magnetic equator at an apex radius of $10\,R$ for three different values of initial Lorentz factor $\gamma_0$. {\color{black} Note that the times here refer to the dimensionaless times (scaled by the respective $t_\mathrm{a}, \S\ref{subsec:energy_loss_rate}$).}}
    \label{fig:de_vs_t_varying_gamma}
\end{figure}

Because the actual minimum value of pitch angle is larger than that considered in the analytical analysis, we also find that the value of $\langle pb^3\rangle_\mathrm{max}$ is smaller than that predicted by Eq. \ref{eq:pb3maxra5}, which can also be seen from Figure \ref{fig:pb3_max_vs_ra}. The consequence is that the value of $r_\mathrm{a}$ beyond which gyrocooling becomes significant is larger than that estimated from Eq. \ref{eq:rabR1}. For example, the numerical values can be reasonably approximated as $0.8\times(r_\mathrm{a}/R)^5$, in which case, the minimum $r_\mathrm{a}$ will be $6.25\,R$ (instead of 5 $R$, \S\ref{subsec:alpha_dep_analytical}). Note, however, that beyond this radius, gyrocooling will dominate over collisional cooling only for the nearly field aligned electrons. Thus, for an isotropic distribution, collisional cooling will still dominate unless we consider very high values of the Lorentz factor. The effect of increasing the initial electron energy ($\gamma_0$) is demonstrated in Figure \ref{fig:de_vs_t_varying_gamma}.

\begin{figure}
    \centering
    \includegraphics[trim={0cm 0cm 0cm 1.5cm} ,clip, width=0.49\textwidth]{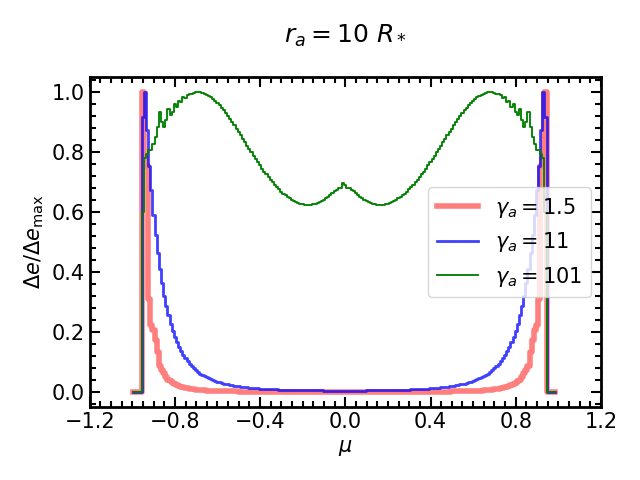}
    \caption{Distribution of energy lost by radiation over co-latitude for a gyrotropic electron distributions with three different values of initial Lorentz factor. The apex radius for all three cases is 10 $R$.}
    \label{fig:de_vs_mu_gyrotropic_dist}
\end{figure}


Finally, for the case of a gyrotropic distribution of electrons injected at an apex radius of $10\,R$, Figure \ref{fig:de_vs_mu_gyrotropic_dist} 
shows that, except for the ultra-relativistic case shown in green,  most of the energy deposition occurs close to the magnetic poles.

\subsection{Spatial distribution of emission from Gaussian deposition}\label{subsec:varying_ra_mu}

\begin{figure*}
\centering
    \includegraphics[width=0.3\textwidth]{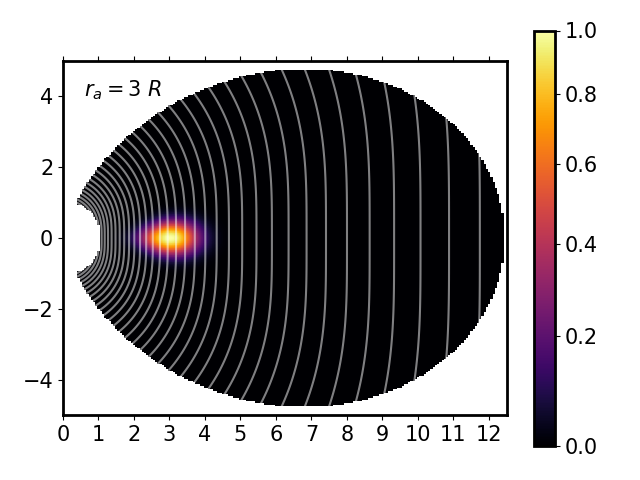}
    \includegraphics[width=0.3\textwidth]{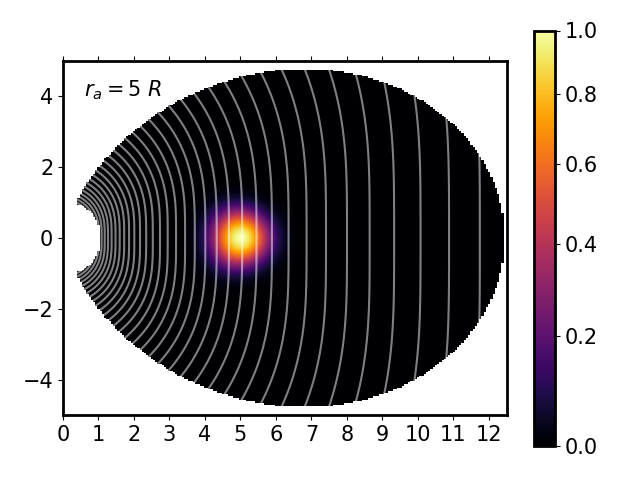}
    \includegraphics[width=0.3\textwidth]{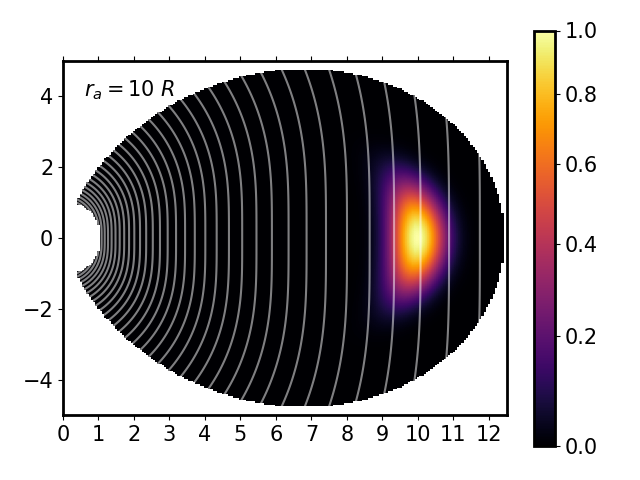}
    \includegraphics[width=0.3\textwidth]{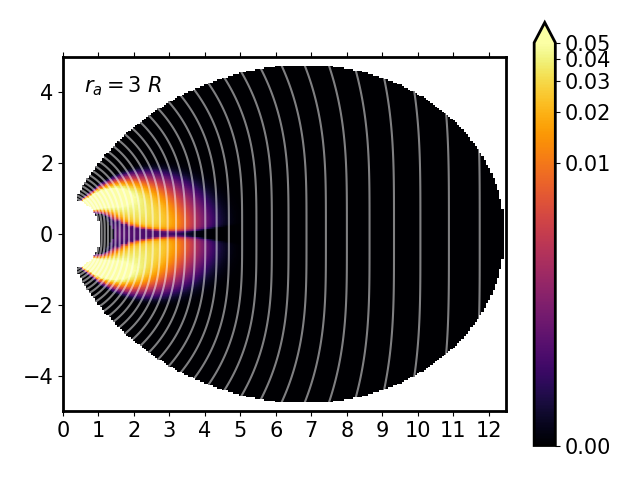}
    \includegraphics[width=0.3\textwidth]{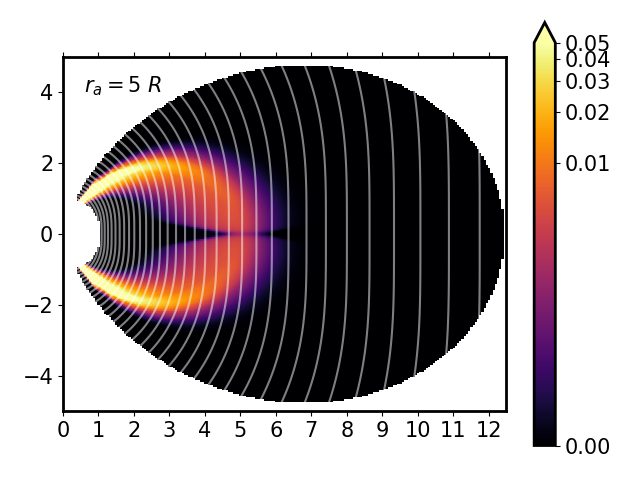}
    \includegraphics[width=0.3\textwidth]{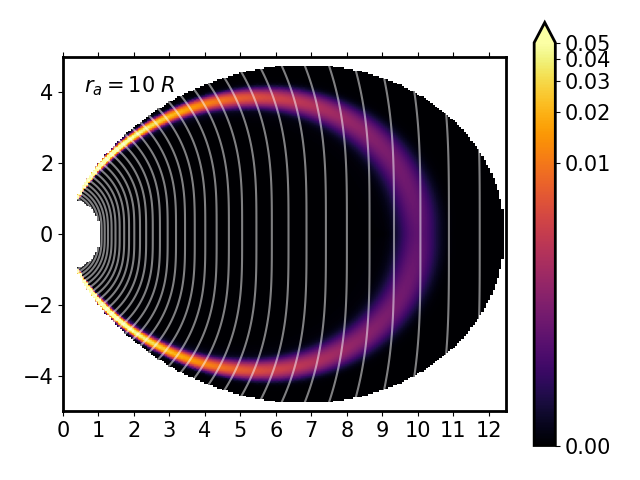}
    \includegraphics[width=0.3\textwidth]{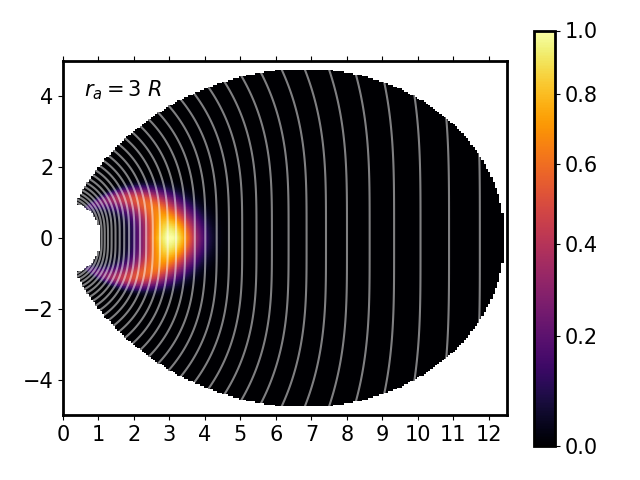}
    \includegraphics[width=0.3\textwidth]{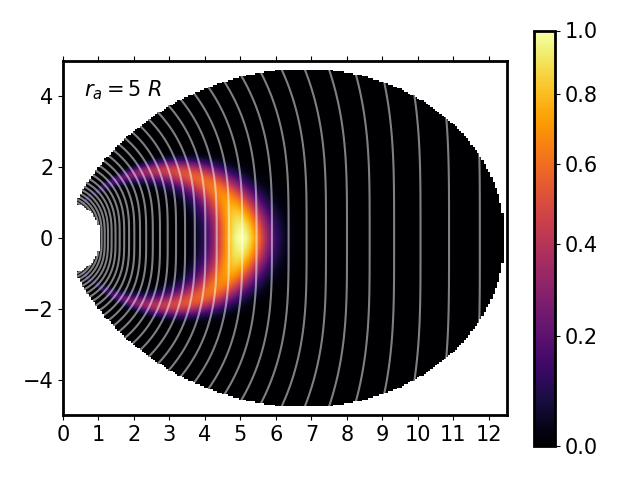}
    \includegraphics[width=0.3\textwidth]{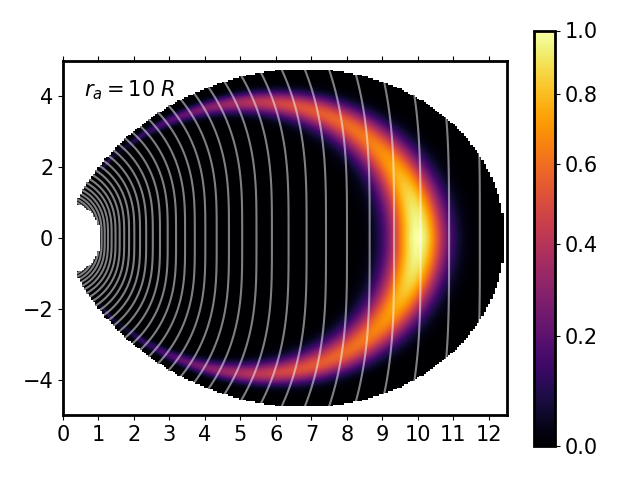}
    \caption{{\color{black}\textit{Top:} Spatial distribution of input energy (Equation \ref{eq:dedrdmu}) for three different values of mean $r_\mathrm{a}$ . The mean Lorentz factor and the parameter $\sigma_\mu$ (see \S\ref{subsec:varying_ra_mu}) are fixed at $1.5$ and $0.1$ respectively. The Kepler radius is $1.59\,R$. \textit{Middle:} The corresponding distribution of energy lost via radiation. The white lines represent contours of magnetic field strength B, spaced logarithmically by $-0.1$ dex from the stellar surface value at the magnetic equator.
    For comparison, the corresponding distributions of radiative energy lost in the absence of collisional cooling are also shown in the bottom panels.
    Note that, since we are interested in the distribution only, we have normalised the energy values by dividing them by the respective maxima (which vary significantly between the cases with and without collisional cooling, but are comparable for the different values of mean $r_\mathrm{a}$).
    }\label{fig:2D_dist_varying_mean_ra}}
\end{figure*}

\begin{figure*}
    \includegraphics[width=0.8\textwidth]{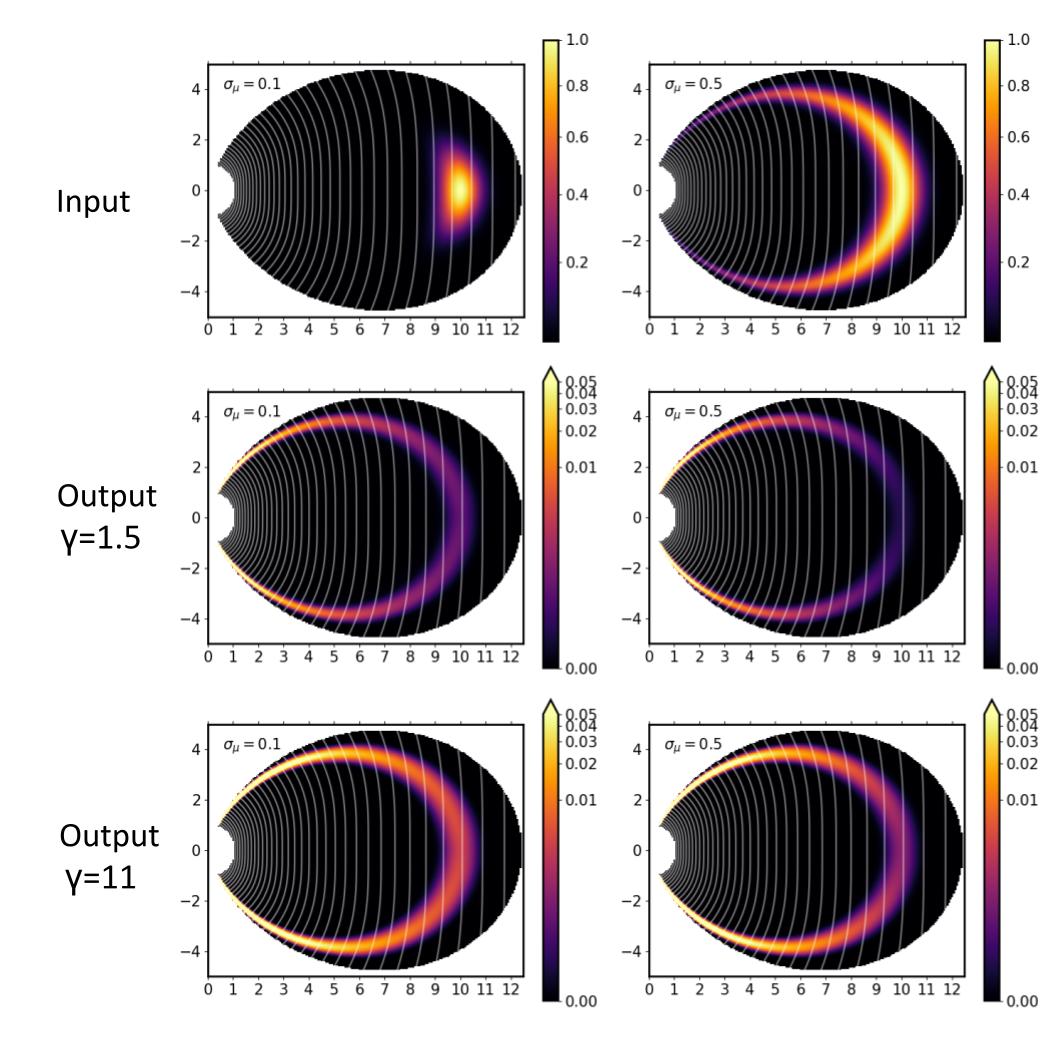}
    \caption{\textit{Top:} Spatial distribution of input energy (Equation \ref{eq:dedrdmu}) for two values of $\sigma_\mu$. \textit{Middle and bottom:} The corresponding distribution of energy lost via radiation for a mean Lorentz factor of 1.5 (middle row) and 11 (bottom row). Note that, since we are interested in the distribution only, we have normalised the values by dividing them by the respective maximum values. The maxima of the output energy distributions are higher for the case of the higher Lorentz factor by approximately an order of magnitude. For a fixed value of the Lorentz factor, more energy is lost via radiation when $\sigma_\mu$ is higher. The white lines again represent contours of magnetic field strength B, spaced logarithmically by $-0.1$ dex from the stellar surface value at the magnetic equator.\label{fig:2D_dist}}
\end{figure*}

To complete comparisons with Paper I, let us next  consider a model in which the initial energy deposition has a gaussian spread in both field co-latitude $\mu=\sqrt{1-r/r_\mathrm{a}}$ and apex radius $r_\mathrm{a} = r/(1-\mu^2)$, centred on a peak radius $r_p$ with radial dispersion $\sigma_{r}$ (cf.\ Equation 16 of Paper I),
\beq
e_{\mu,r_\mathrm{a}} =
C_{\mu,r_\mathrm{a}}\exp\left[ -\frac{{(r_\mathrm{a}-r_p)}^2}{2\sigma_{r}^2}\right] \exp\left(-\frac{\mu^2}{2\sigma_\mu^2}\right)
\, ,
\label{eq:dedrdmu}
\eeq

where $C_{\mu,r_\mathrm{a}}$ is a normalization factor such that 
\begin{align*}
    \int_{R}^{\infty} \int_{-\mu_\ast}^{+\mu_\ast} e_{\mu,r_\mathrm{a}}d\mu\,dr_\mathrm{a}&=1
    \, ,
\end{align*}
with $\mu_\ast \equiv \sqrt{1-R/r_\mathrm{a}}$.

Figure \ref{fig:2D_dist_varying_mean_ra} shows the distribution of radiative energy loss for mildly relativistic electrons ($\gamma_0=1.5$) with $\sigma_\mu=0.1$ and $\sigma_r=0.5$ deposited around three different values of the apex radius: $3\,R$ (left), $5\,R$ (middle) and $10\,R$ (right). The top panel shows the input energy distribution, the middle panel shows the resulting radiative cooling taking into account of collisional cooling (this work) and the bottom panel shows the corresponding distribution in the absence of thermal plasma causing collisional cooling (Paper I). Note that the net emission is scaled by the respective maxima, which differ significantly, since, the total radiative energy lost is significantly smaller (by $\approx 2$ orders of magnitude) when collision is considered for this mildly relativistic case (e.g. see the top panel of Figure \ref{fig:de_vs_t_varying_gamma}). In terms of spatial distribution of emission, the key difference is that, in the absence of collisional cooling, radiative cooling predominantly occurs close to the magnetic equator, but when collisional cooling is taken into account, the high-density plasma at the magnetic equator suppresses radiative loss, especially for smaller values of mean $r_\mathrm{a}$, and the electrons lose their energy via radiation predominantly close to the magnetic poles only. Also, for the pure radiative cooling case (Paper I), the spatial distribution remains qualitatively the same for the different values of the mean deposition radius. However, in the presence of collisional cooling, the radiative energy loss close to the magnetic equator gradually increases with increasing apex radius. This is because the thermal plasma density falls sharply away from the Kepler radius (taken to be $1.59\,R$, Figure \ref{fig:CBOCM}).

Finally, we examine the effect of varying the electron energies, as well as the distribution of the initial energy deposition.
For cases with {\color{black}$r_\mathrm{a} = 10\,R$} and $\sigma_r=0.5\,R$, the top row of Figure  \ref{fig:2D_dist} shows the assumed energy deposition $e_{\mu,r_\mathrm{a}}$ for cases with $\sigma_\mu = 0.1$ (left column) and $\sigma_\mu=0.5$ (right column).
The rows below show the resulting net emission (scaled here by their respective maxima) for cases with moderately relativistic ($\gamma_0=1.5$; middle row) and highly relativistic ($\gamma_0=11$; bottom row) injection energies.

The results show that, rendered and scaled in this way, the computed spatial distribution of emission is surprisingly independent of the assumed
parameter for deposition width $\sigma_\mu$ and even deposition energy $\gamma_0$.
In all cases, the dominant emission occurs near the loop footpoint, where the field strength is strongest, leading to strong gyro-emission,
with also minimal collisional cooling, due to the large distance from the dense CM region around the loop top.
The more highly relativistic case does also show a relatively greater level of such scaled emission near the  dense loop top, owing to the reduced collisional cross section at for higher electron energies.

\subsection{Emission distribution as a function of magnetic field strength}\label{subsec:dl_db}

The great distance of massive stars makes it difficult to resolve directly the spatial distribution of radio emission from their magnetospheres.
{\color{black}But because the electron gyro-frequency scales} directly with magnetic field strength, characterising the emission distribution with field strength
provides an initial basis for deriving observable gyro-emission spectra.

\begin{figure}
    \centering
    \includegraphics[width=0.49\textwidth]{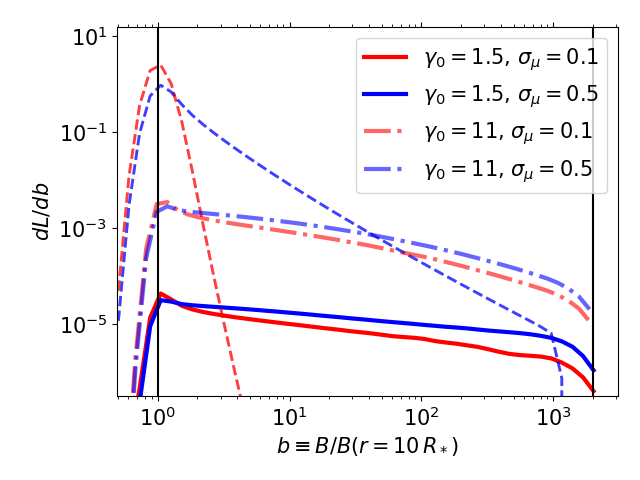}
    \caption{Distribution of radiated energy as a function of magnetic field strength corresponding to the cases shown in Figure \ref{fig:2D_dist}. The dashed lines represent the input energy distribution. The vertical line on the right represents the polar magnetic field strength.}
    \label{fig:spectrum_2D}
\end{figure}

Let us thus define a differential emission luminosity $dL \equiv \epsilon_{\mu,r_\mathrm{a}}d\mu dr_\mathrm{a}$ associated with a spatial differential $dr_\mathrm{a}$ of magnetic loops around a central apex $r_\mathrm{a}$, and a co-latitude differential $d\mu$ along the loops.
For a dipole field, we can readily identify the associated differential in field strength $db$, scaled by the loop strength at the central loop apex $r_\mathrm{a}$.
For the same set of parameters shown in Figure \ref{fig:2D_dist}, Figure \ref{fig:spectrum_2D} now plots the  emission distributions $dL/db$ versus the scaled field $b$.
The dashed curves show the corresponding input distributions for $\sigma_\mu = 0.1$ (red) and $\sigma_\mu = 0.5$ (blue).
The other associated curves show results for the cases $\gamma_0=1.5$ (solid) and $\gamma_0=11$ (dot-dashed).

Reflecting the much reduced collisional loss of energy for higher energy electrons, the curves for the strongly relativistic case $\gamma_0=11$ (dot-dashed curves) are about two orders of magnitude higher than those for the mildly relativist case $\gamma_0=1.5$ (solid curves).

But the similarly broad distribution of the emission for the red versus blue curves, for spatially narrow ($\sigma_\mu=0.1$) versus broad ($\sigma_\mu=0.5$) energy deposition along the loops, again demonstrates the surprising insensitivity of the resulting emission to assumptions about this spatial distribution of energy deposition.

Incoherent radio emission from magnetic hot stars is interpreted as gyrosynchrotron emission \citep[e.g.][]{drake1987} for which the emission occurs at the harmonics of the electron gyrofrequency that is proportional to the local magnetic field strength.
Because of the complexity of the emission spectrum (for a fixed magnetic field) and the requirement to perform detailed radiative transfer modeling, quantitative derivation of observed spectra is beyond the scope of this paper,  so will be deferred future work.
But the overall results here suggest that such spectra should generally have a broad distribution in frequency.

\section{Discussion}\label{sec:discussion}
\subsection{Potential application to magnetic ultra-cool dwarfs}\label{subsec:BDs}
Similar to the magnetic hot stars, large-scale (dipolar), kG-strength magnetic fields have also been observed in ultracool dwarfs (UCDs), including brown dwarfs. Despite their much lower surface temperature,
their magnetospheric radio emission exhibits striking similarities with that of magnetic hot stars.
Both types of objects are known to be capable of producing periodic radio pulses by electron cyclotron maser emission \citep[][etc.]{trigilio2000,berger2001}. 
\citet{leto2021} showed that an empirical relation connecting incoherent radio luminosity and magnetospheric parameters derived for magnetic hot stars could also hold in the UCD regime, and even for Jupiter. Most recently, a brown dwarf's magnetosphere was resolved for the first time using Very Long Baseline Interferometry \citep[VLBI,][]{kao2023,climent2023}.
These images revealed that the radio emission is highly extended and has a double-lobed morphology distributed in the magnetic equatorial plane similar to the radiation belt observed in solar system planets. \citet{kao2023} also inferred a Lorentz factor of $\gamma\sim 30$ and speculated that the CBO mechanism, proposed for magnetic hot stars, might also be in play in the magnetospheres of brown dwarfs.

Both the locations of the radio emitting sites (magnetic equator) and the presence of highly relativistic electrons contradict the scenario for magnetic hot stars. The properties of incoherent radio emission from magnetic hot stars are consistent with gyrosynchrotron suggesting that the non-thermal electrons are mildly relativistic \citep[][etc.]{drake1987,linsky1992,trigilio2004}. The rotational modulation of incoherent radio flux density and their circular polarisation suggest that the radio emitting cites are located along the magnetic poles \citep[][etc.]{leone1993,leto2021}. Thus, although the magnetospheres of hot stars are yet to be resolved, the current data appear to suggest that the magnetospheres of hot stars and UCDs have certain fundamental differences.

The formalism presented in this work provides a plausible explanation for this difference. A key characteristic of hot stars that separates them from the UCDs (and other cool stars) is that the former drives a strong stellar wind, but the latter cannot. A consequence could be that brown dwarfs may not have sufficient thermal plasma in their magnetospheres to cause collisional cooling of non-thermal electrons. In addition, the high Lorentz factor ($\sim 30$) of the electrons will further suppress collisional cooling. From Figure \ref{fig:2D_dist_varying_mean_ra}, we find that in the absence of collision, radiative cooling of electrons predominantly occurs around the magnetic equator. The introduction of thermal plasma at the equator increases collisional cooling there, so the regions close to the magnetic poles become the primary sites of radiative energy loss. Thus, it is possible that the relative importance of collisional cooling is responsible for the difference in the spatial radio maps of magnetic hot stars and magnetic UCDs\footnote{On the other hand, if the acceleration of high-energy electrons in UCDs is taken to be from 
CBO-induced magnetic reconnection, the density of thermal plasma should follow the scalings assumed here, which are insensitive to whatever wind or other mechanism that is populating the magnetosphere.}. 
The reason behind the difference in the energetic electron population is, however, yet to be understood.

\subsection{VLBI imaging of magnetic hot stars}\label{subsec:VLBI_HMS}
Recently, \citet{klement2025} reported near-infrared spectrointerferometric observation of a magnetic hot star $\rho$ Oph A. These observations enabled them to detect changes in the location of the emission regions as a function of stellar rotational phases and verify that the locations are consistent with those predicted from the model of H$\alpha$ emission. Similarly, to test existing notions about the sites of radio emission and the emission morphology, it is important to conduct VLBI experiments at radio bands on magnetic hot stars. The best angular resolution that can be obtained with the current generation of VLBI instruments at a representative frequency of 8 GHz is $\sim 1$ mas\footnote{{\color{black}e.g. \url{https://evlbi.org/capabilities}}}. 
Following the strategy of \citet{kao2023}, the angular extent of the magnetosphere of a star with a dipolar magnetic field of polar surface strength $B_{\rm kG}$ (in kG) can be estimated using the following relation:

\begin{align}
    \theta_{\nu,s}&= \frac{2R}{d}{\left(\frac{B_\mathrm{kG}\times 2.8 s}{\nu_{\rm GHzx}}\right)}^{1/3} ,
    \label{eq:mag_size}
\end{align}
where $s$ is the harmonic number, $d$ is the distance to the star, and $\nu_{\rm GHz}$ is the frequency of observation in GHz.
$B_\mathrm{kG}\times 2.8$ gives the electron gyrofrequency in GHz corresponding to a magnetic field strength of $B_\mathrm{kG}$. The above formula assumes that the radio emitting sites are located along the magnetic poles and $\theta_{\nu,s}$ is the separation between the radio emitting sites located in opposite magnetic hemispheres (which introduces the factor of `2').
For gyrosynchrotron emission, $10\lesssim s \lesssim 100$ \citep{gudel2002}.

\begin{figure}
    \centering
    \includegraphics[width=0.49\textwidth]{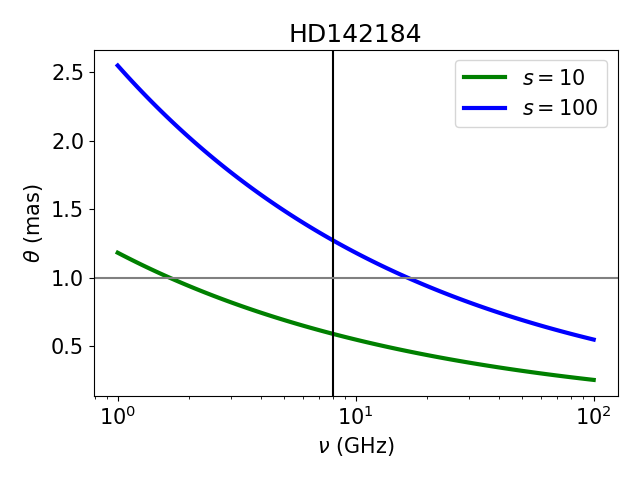}
    \caption{Variation of the angular extent of the magnetosphere of the radio-bright magnetic hot star HD\,142184, as a function of frequency for different harmonic numbers $s$, as predicted by Equation \ref{eq:mag_size}. The stellar and magnetic parameters are taken from \citet{shultz2019c} and the distance is obtained from the Gaia parallax measurement \citep{gaia2016,gaia2023}. The vertical line marks the frequency of 8 GHz, and the horizontal line marks $\theta=1$ mas.}
    \label{fig:theta_vs_freq}
\end{figure}

The radio-bright magnetic hot star HD\,142184 (also known as HR\,5907) has the highest incoherent radio luminosity in the sample of \citet{leto2021}. It has a strong polar magnetic field strength of $B_{\rm kG} \approx 9$ \citep{shultz2019c} and is located at a moderate distance of $143$ pc \citep{gaia2016,gaia2023}.
The angular size of its magnetosphere as a function of frequency is shown in Figure \ref{fig:theta_vs_freq} for $s=10$ and $s=100$, which approximately mark the lower and upper limits of the angular extents, respectively. Thus, in principle, the star's magnetosphere can be resolved with existing VLBI instruments. Note, however, that the star has a declination of $\approx -24^\circ$, which is not ideal for achieving the best performance from Northern VLBI instruments like the European VLBI Network (EVN) and the High Sensitivity Array (HSA).

\section{Summary}\label{sec:summary}
This paper examines the effect of high density plasma on the spatial distribution of radio emission produced by large-scale stellar magnetospheres. 
This is an extension of the work presented in Paper I that provides a framework to calculate such spatial distribution, albeit without taking into account collisional loss.
For simplicity, we have assumed an axisymmetric dipolar case aligned to the rotation axis such that CBOs take place close to the magnetic equatorial plane. This justifies the assumed scenario of non-thermal electron deposition symmetrically around the magnetic equator. 

The key result from \citet{Das2023} is that for a low-energy electron distribution, and when the energy is injected over a relatively small spatial scale, the emission occurs primarily at the magnetic equatorial plane, but as the energy deposition occurs over a larger spatial scale, the emission sites get shifted towards the regions with stronger magnetic fields, i.e., closer to the stellar surface. They also found that increasing the energy of the electron distribution has only a modest effect on the spatial distribution.

The addition of collision significantly changes the above scenario. Since for the CBO model, the high density plasma resides primarily around the Kepler radius and around the magnetic equator, radiative loss is strongly suppressed close to the magnetic equator and the emission sites are pushed close to the magnetic poles. At the same time, radiative loss is more efficient away from the Kepler radius. Thus, radiative loss occurs along magnetic field lines with equatorial radii significantly higher than the Kepler radius, and at sites close to the magnetic poles. This is consistent with the currently assumed picture of hot magnetic stars' radio emission. 

The main takeaways from this work are listed below:
\begin{enumerate}
    \item Even though CBOs can occur all over the CM (around the magnetic equator for {\color{black}an} aligned rotator) depositing non-thermal electrons, radio emission is more efficient along magnetic field lines lying in the outer parts of the CM away from the Kepler radius.
    \item The emission sites are most likely located along the magnetic axis. Note that this was already suggested based on the correlation observed between the rotational modulation of incoherent radio emission as well as the circular polarisation fraction with that of the longitudinal magnetic field \citep[e.g.][]{linsky1992, leone1993}. 
    While our work only considers the case of aligned rotators, for which no rotational modulation in either the longitudinal magnetic field or the incoherent radio flux density is expected, it provides an explanation for the inferred spatial distribution of sites under the assumption that the effect of a tilted magnetic pole is not significant in this context. This, however, need not be true for highly misaligned rotators (rotation and dipole axes are misaligned by $\sim 90^\circ$) as the corresponding thermal plasma density is predicted to be significantly different from that for an aligned rotator \citep{townsend2005,
     ud-doula2023}. 
    \item The location of emission sites in hot stars' magnetospheres, thus appear to be different from that for magnetic {\color{black} ultra-cool} dwarfs, where the emission sites are believed to lie at the magnetic equatorial plane \citep{kao2023, climent2023}. However, this could be due to the fact that in brown dwarf's case, collisional cooling does not play any significant role.
    \item The effect of collisional cooling can be compensated for by considering ultra-relativistic electrons (e.g. Lorentz factor {\color{black}$\gamma > 10$)}. However, for mildly relativistic to relativistic electrons ($\gamma\sim 1-10$), collision plays a very strong role in determining the spatial distribution of emission. 
    Note that electrons in hot stars' magnetospheres are believed to be mildly relativistic with $\gamma\gtrsim 1.2$ \citep{trigilio2004}, so that one cannot ignore the role of collisional cooling. Interestingly, for brown dwarfs, a recent study has suggested the presence of relativistic electrons with $\gamma\sim 30$ \citep{kao2023}.
    \item The emission distribution is only slightly sensitive to the distribution of energy deposition along the co-latitude axis.
\end{enumerate}

Both in Paper I and this work, it is assumed that the initial non-thermal electrons have a gyrotropic distribution. This is essentially a simplifying assumption, the actual parent electron distribution can have pitch angle anisotropy ingrained during the magnetic reconnection process \citep[e.g.][]{comisso2023}. A more field aligned distribution will lead to enhanced emission closer to the stellar surface compared to that at the apex radius.

While the current framework only provides a qualitative understanding of the spatial distribution, the results will allow us to obtain insights about the energy deposition by CBOs if the spatial distribution of emission can be actually observed. 
As outlined in \S\ref{subsec:VLBI_HMS}, the magnetic hot stars can have angular extents that are, in principle, resolvable with existing VLBI instruments. 
The growing number of radio-bright magnetic hot stars in recent times \citep[][etc.]{das2025,driessen2024} should further enhance the feasibility of these experiments. The ultimate goal will be to obtain rotational phase resolved spatial radio maps spanning a wide frequency range. This will allow testing not only the relative importance of collisional cooling, but also the current notions about magnetospheres of hot stars and their similarities (differences) with those of brown dwarfs.
 

\paragraph{Acknowledgement}
We thank the referee for their constructive criticism that has helped us significantly improve the work.    
This work has made use of data from the European Space Agency (ESA) mission
{\it Gaia} (\url{https://www.cosmos.esa.int/gaia}), processed by the {\it Gaia}
Data Processing and Analysis Consortium (DPAC,
\url{https://www.cosmos.esa.int/web/gaia/dpac/consortium}). Funding for the DPAC
has been provided by national institutions, in particular the institutions
participating in the {\it Gaia} Multilateral Agreement.

\paragraph{Funding Statement}
The contributions by SPO are supported in part by the National Aeronautics and Space Administration under Grant No. 80NSSC22K0628 issued through the Astrophysics Theory Program.
\paragraph{Competing Interests}
None.

\paragraph{Data Availability Statement}
This is a theoretical work, and does not directly use any observational data.

\bibliographystyle{paslike}
\bibliography{main}

\end{document}